\def\etal{{\it et al. }}

 \documentclass[final,3p,times,twocolumn]{elsarticle}
\usepackage{amssymb}
\usepackage{amsmath}
\usepackage{float}
\usepackage{subfigure}
\usepackage{color}
\usepackage{adjustbox}
\usepackage{color, colortbl}
\definecolor{Gray}{gray}{0.75}
\definecolor{LightCyan}{rgb}{0.50,1,1}

\usepackage[section]{placeins}
\usepackage{amsfonts}
\usepackage{graphicx}
\usepackage{lipsum}
\usepackage{multirow}
\usepackage{booktabs}
\usepackage[table,xcdraw]{xcolor}
\definecolor{cream}{RGB}{222,217,201}

\journal{Acta Materialia}

\begin {document}
\begin{frontmatter}

\title{First-principles prediction of incipient order in arbitrary high-entropy alloys:  exemplified in Ti$_{0.25}$CrFeNiAl$_{x}$}

\author{Prashant Singh}
\address{Ames Laboratory, U.S. Department of Energy, Iowa State University,  Ames, Iowa 50011, USA}
\author{A.V. Smirnov}
\address{Ames Laboratory, U.S. Department of Energy, Iowa State University,  Ames, Iowa 50011, USA}
\author{Aftab Alam} 
\address{Department of Physics, Indian Institute of Technology, Bombay, Powai, Mumbai 400076, India}
\author{Duane D. Johnson}
\address{Ames Laboratory, U.S. Department of Energy, Iowa State University,  Ames, Iowa 50011, USA}
\address{Materials Science \& Engineering, Iowa State University, Ames, Iowa 50011, USA}

\begin{abstract}

Multi-principal-element alloys, including high-entropy alloys, experience segregation or partially-ordering as they are cooled to lower temperatures. For Ti$_{0.25}$CrFeNiAl$_{x}$, experiments suggest a partially-ordered B2 phase, whereas CALculation of PHAse Diagrams (CALPHAD) predicts a region of  L2$_{1}$+B2 coexistence. We employ first-principles density-functional theory (DFT) based electronic-structure approach to assess stability of phases of  alloys with arbitrary compositions and Bravais lattices (A1/A2/A3). In addition, DFT-based linear-response theory has been utilized to predict Warren-Cowley short-range order (SRO) in these alloys, which reveals potentially competing long-range ordered phases. The resulting SRO is uniquely analyzed using concentration-waves analysis for occupation probabilities in partially-ordered states, which is then  be assessed for phase stability by direct DFT calculations. Our results are in good agreement with experiments and CALPHAD in Al-poor regions ($x \le 0.75$) and with CALPHAD in Al-rich region ($0.75 \le {x} \le 1$), and they suggest more careful experiments in Al-rich region are needed. Our DFT-based electronic-structure and SRO predictions supported by concentration-wave analysis are shown to be a powerful method for  fast assessment of competing phases and their stability in multi-principal-element alloys.
\end{abstract}


\end{frontmatter}

%(site-occupation probabilities) 

\section{Introduction}
Multi-principal-element alloys or complex solid-solution alloys, of which high-entropy alloys (HEAs) are a subset, have established a new paradigm in alloy design \cite{Yeh2004}, but there are many  fundamental science questions unanswered. These HEAs offer a huge unexplored composition space that has lead to the discovery of novel alloys with unusual properties \cite{Yeh2004,MS2017,Huang1996,Yeh2006,Shigenobu2009,Otto2013,M2019,George2019,Ikeda2019,Murty2014,Gao2016,Gao2013}. Short-range order (SRO) is one such key property that remains less explored in HEAs \cite{GS1983,GJPNS1989,SJP1990,AJP1995,DDJ2012,SSJ2015,KSS2016}. SRO in disorder phase is indicative of  the  expected partially or fully-ordered state from an order-disorder transition, where two distant atoms are connected by non-zero correlations as described by long-range order (LRO) parameters \cite{Krivoglaz1969, Clapp1966, Ducastelle1991,Khat1972, Khat1983,Fontaine1975,Fontaine1979}. The site-pairwise correlation between atoms in an alloy is given by the Warren-Cowley SRO parameters \cite{SSJ2015}. {In terms of diffraction, the} SRO is usually defined with respect to the {underlying (average) crystal lattice in the high-temperature phase, i.e., FCC, BCC, or HCP, and competes within the alloy as temperature is decreased}. By controlling SRO and LRO, the processing and properties of advanced materials can be manipulated \cite{NiCoCr2019}. Thus, predicting SRO in HEAs and assessing its electronic origins (e.g., band-filling, Fermi-surface, atomic-size [i.e., band-width], or charge-transfer)  is of great importance.

{\par} {The SRO, in principle, can be determined experimentally from diffuse-scattering intensities measured in reciprocal space using x-ray, neutron, or electron diffraction \cite{Sato1962, Moss1969, Reinhard1990}, which depends on the differences of atomic scattering factors. However, similar structure factors of some alloying elements in a HEA makes the SRO measurements difficult. Even after these complex measurements, it is not possible to pinpoint the underlying origin of SRO.} Hence, the calculation of diffuse intensities in HEAs based on electronic density-functional theory (DFT) and the subsequent connection of those intensities directly to its origin(s) can provide a fundamental understanding of the experimental data and phase instabilities \cite{DDJ2012}.

{\par} Generally, {the SRO assessed in high-temperature phases (BCC, FCC, and HCP) is not biased by \emph{a priori} choice of LRO phases}, which may be a problem with more traditional total-energy methods \cite{Johnson2000, Zarkevich2004}. Singh {\it et.~al.}~\cite{SSJ2015,Singh2018,Singh2018_2,Singh2019} presented a DFT based thermodynamic linear-response theory that predicts all SRO pairs simultaneously (as done for displacement modes, i.e. phonons) and details the ordering behavior in HEAs \cite{SSJ2015}. {The SRO theory can predict the eigenvectors (chemical modes) at the order-disorder transition with respect to the formation of short-wavelength concentration waves \cite{Khat1972,Khat1983}. These eigenvectors can be used to characterize the potential ordered structures, e.g., B2 phase in BCC alloys, with site-probabilities modulated in a wave-like periodicity \cite{Khat1972,Khat1983,Fontaine1975,Fontaine1979}, all determined from the underlying electronic structure of each specific HEA composition.}

{\par} {In this paper, we explore the phase stability over full Al-composition range in Ti$_{0.25}$CrFeNiAl$_{x}$ HEA with focus on few special compositions, e.g., x=0.25, 0.50, 0.75 and 1.0, allowing one-on-one comparison with available experiments. The formation enthalpy (E$_{form}$) shows that Al stabilizes BCC phase up to three Al mole-fractions and FCC phase at higher Al compositions. For $x \le 0.75$ Al-composition range, our SRO assisted CW predictions of ordering phases compare well with the CALPHAD and experiments of Liu  {\it et.al.}~\cite{L2015}. Interestingly, for Al-composition range $0.75 - 1$,  experiments do not find any secondary ordering, whereas CALPHAD indicates competing B2 and L2$_{1}$ phases. Our CW analysis when combined with direct DFT calculations shows thermodynamically stable B2 and L2$_{1}$ phases. The DFT-based SRO theory combined with CW approach allows a quick assessment of ordering behavior in HEAs (based on the composition and associated electronic-structure of the specific alloy), similar to the Hume-Rothery empirical relationships of phase formation in traditional alloys.}

%Specifically, we

\section{Computational method}
{\par} {Here, we use the first-principles DFT in combination with linear-response theory and analytically formulated concentration-wave (CW) analysis \cite{Khat1983} to study phase-stability, electronic-structure, and ordering behavior of HEAs, as well as identifying the electronic origins.} 

{\it DFT Calculation:} KKR-CPA is a DFT based Green's function electronic-structure approach that permits charge self-consistency and configurational averaging to be done simultaneously (averaged Green's functions are related directly to observables \cite{Martin1968}). The coherent potential approximation (CPA) is used to handle chemical disorder and its configurational averaging,  {and associated Friedel screening} \cite{JohnsonCPA, JP1993,MECCA}. We used generalized gradient approximation (PBE) as the exchange-correlation functional \cite{libxc2012}. The core electrons and semi-core/valence electrons are treated relativistically and scalar-relativistically, respectively. A variational potential zero ${v}_0$ is used to yield kinetic energies and dispersions nearing those of full-potential methods \cite{AJ2012}. 
%The Atomic Sphere Approximation (ASA) represents each Voronoi polyhedra (VP) in a unit cell but with periodic boundary conditions enforced and VP integrations are used for charge distributions. 
For self-consistent densities, a 20 (complex) energy point Gauss-Legendre semi-circular contour integration is used  {to integrate the Green's function} \cite{J1985}, and $L_{max}(l,m)$=3 in spherical-harmonic basis.
Brillouin zone (BZ) is sampled using Monkhorst-Pack method \cite{Monkhorst} with $12\times12\times12(6)$ for FCC, BCC (HCP) meshes. 
We used 300 $\it{k}$-points along symmetry lines to visualize electronic dispersion, {i.e., ``band'' structure}. 

{\par}{\it Chemical SRO:} Formally, the Warren-Cowley SRO parameters (pair correlations), i.e.,  $\alpha_{\mu\nu} ({\bf k};T)$ in Laue units for atom pairs $\mu$,$\nu$, are defined relative to average x-ray scattering lattice, as atomic displacements sum to zero on average (by symmetry, for each spatial direction). For linear-response, the second-order variation of  DFT free-energy with respect to elemental concentrations $\{c^{i}_{\mu}\}$ at sites $i$,$j$ is performed \cite{SJP1994,AJP1995,DDJ2012,SSJ2015}, and the Warren-Cowley SRO are analytically found for any N-component HEA with concentrations $\{c_{\mu}\}$ at temperature (T)  to be given by an $({N}-1) \times ({N}-1)$ matrix 
\begin{eqnarray} \label{ASRO}
\small{
[\alpha^{-1}({\bf k};T)]_{\mu\nu} =  {\cal{C}}_{\mu\nu}    - \beta  c_{\mu}(\delta_{\mu\nu} - c_{\nu}) S_{\mu\nu}^{(2)}({\bf k}; {T}) }  \\
{ \text{where}} \,\,\,\,  {\cal{C}}_{\mu\nu} = c_{\mu}(\delta_{\mu\nu} - c_{\nu})  { (\frac{\delta_{\mu\nu}}{c_{\mu}} + \frac{1}{c_N}) }  
  \nonumber
\end{eqnarray}
where $\beta^{-1} = k_B{T}$  with $k_B$ is the Boltzmann constant and ${\cal{C}}_{\mu\nu}$ is a  constant matrix element \cite{SSJ2015}. The arbitrary N$^{th}$ atom is used as ``host'' due to conservation of atoms ($\text{N}-1$ independent concentrations, i.e., $\sum_{\mu=1}^{\text{N}} c_{\mu}=1$). The pairwise-interchange energy S$_{{\mu\nu}}^{(2)} ({\bf k}; {T})$ in linear-response is the \emph{chemical stability matrix}  referenced to the homogeneous HEA \cite{DDJ2012}, which reflects the free-energy cost for all pair fluctuations with $\{c^{i}_{\mu}({\bf k})\}$ \cite{SSJ2015,Singh2018,Singh2018_2,Singh2019}. {Here, S$_{{\mu\nu}}^{(2)}({\bf k}; {T})$ is the energy cost to exchange specific atomic pairs, with effects included to all orders in the electronic structure. S$_{{\mu\nu}}^{(2)} ({\bf k}; {T})$ also reveals the unstable ordering modes, its origin as well as the ``fingerprint'' for the ordering behavior in an arbitrary HEA \cite{SJP1990,SSJ2015,AJP1995,DDJ2012,AS2016,Pinski1991,AJPS1996,SJP1994}.}

{\par} For completeness, we note that Eq.~\ref{ASRO} is exact \cite{DDJ2012,Evans1979}, if the configurational average of the functional and its variation are handled exactly. However, the CPA is a mean-field (single-site) approximation to the average, and, although it is often adequate, it is not exact. Nonetheless, while it can be made increasingly accurate via  a cluster generalization of the CPA \cite{Jarrell1998}, a simple Onsager correction to the single-site CPA is sufficient to remove most of the error \cite{DDJ2012}. Moreover, this correction eliminates the incorrect topology of mean-field phase diagrams and dramatically improves the transition temperatures \cite{TTJ2011}. 
{In short, the mean-field S$^{(cpa)}_{\mu\nu}$ is corrected by requiring that the SRO intensity (Eq.~\ref{ASRO}) properly conserves the sum rule, i.e., $\alpha^{ii}_{\mu\nu} = 1- {\delta_{\mu\nu}}/{c_{\beta{i}}}={V^{-1}_{BZ}} {\int{d\bf{k}}~\alpha_{\mu\nu}(\bf{k})}$ \cite{DDJ2012}. 
The outcome in N-component alloy is that $\alpha_{ii}^{\mu\nu}$ are normalized correctly and the unmeasured correlations (diagonal pairs, i.e., $\alpha_{ii}^{\mu\mu}$) are correctly determined. For more details see~[\citenum{DDJ2012}]~and~[\citenum{TTJ2011}].} 
Minimally, a site-diagonal (${\bf k}$-independent) self-energy ($\Lambda^{ij}_{\mu\nu}=\Lambda^{ii}_{{\mu\nu}}\delta_{ij}$)  is required, i.e., 
\begin{eqnarray} \label{ASRO2}
S_{{\mu\nu}}^{(2)} &\approx&  S_{\mu\nu}^{(cpa)} - \Lambda_{\mu\nu}(T) ,      \\
   \Lambda_{{\mu\nu}}(T) &=& \frac{1}{V_{BZ}} \int d{\bf k}~S_{{\mu\pi}}^{(cpa)} \alpha({\bf k}; \Lambda ; {T})]_{\pi\nu} , \nonumber
\end{eqnarray}
where the SRO is implicitly dependent on $\Lambda_{{\mu\nu}}(T)$, and must be corrected at each ${T} >  {T}_{sp}$, the spinodal temperature (see below). At large temperatures,
$\Lambda_{\mu\nu}(\infty)=0$ and this coupled set of equations can be solved iteratively by Newton-Raphson \cite{SJP1994,AJP1995,SSJ2015}.

{\par} {\it Spinodal Decomposition \& Transition Temperatures:} The most unstable SRO mode is where $\alpha_{\mu\nu} ({\bf k}_{o}; T > T_{sp})$ has the largest peak at ${\bf k}_{o}$ for a specific $\mu$-$\nu$ pair \cite{SSJ2015}. An absolute instability to the ${\bf k}_{o}$ mode occurs below the spinodal temperature T$_{sp}$ defined at $[\alpha^{-1} ({\bf k}_{o};T_{sp})]_{\mu\nu}=0$.  The normal modes are eigenvectors of S$^{(2)}_{\mu\nu} ({\bf k}_{o}, T \rightarrow T_{sp})$ driving divergence in SRO. These modes are obtained for any arbitrary HEA from S$_{{\mu\nu}}^{(2)} ({\bf k}; {T})$ using a special oblique coordinate transform in a given Gibbs space \cite{SSJ2015}. 

{\par} {\it Concentration-Wave Analysis of SRO:} {The idea of concentration-wave (CW) for multi-component alloys is adopted from Khachaturian \etal \cite{Khat1983}, which was found useful in characterizing ordering phases of stoichiometric binary compositions. Since, the present day multi-component alloys are neither stoichiometric nor binary, it increased the complexity of the problem.} With normal modes of eigenvectors from the SRO Eq.~\ref{ASRO}, a vector n$({\bf r})$ of probabilities for each element to occupy specific sites in a crystal structure for partially-LRO cell or superlattice, i.e., the generalized concentration-waves, can be written as
\begin{equation}
n({\bf r}) =  c({\bf r}) +  \sum_{s,\sigma}\eta_{\sigma}^{s} (T) \nu_{\sigma}({\bf k}_{s})\sum_{j_{s}}\gamma_{\sigma}({\bf k}_{j_{s}})e^{i{\bf k}_{j_{s}}\cdot{\bf r}}   .
\label{NCompCW}
\end{equation}. 
Here, ${c}({{\bf r}})$ is an $(N-1)$-component vector in site occupation probabilities $\{c_{\mu}\}$ in the Bravais lattice of the N-component  homogeneous HEA reference. Whereas, ${n}({{\bf r}})$  depends on the type of order and real-space site coordinates, dictated by the LRO parameters $\eta(T)$. The sum $s$ runs over the ``stars'' (inequivalent ${\bf k}$ that define the order), $j_{s}$ (equivalent ${\bf k}_{j_{s}}$ in the $s^{th}$-star), and $\sigma$ (eigenvector branch of the free-energy quadric). The other quantities are LRO parameter $\eta_{s}^{\sigma}(T)$ for the $\sigma^{th}$ branch and $s$ star; $\nu_{\sigma}$ is $(N-1)$-component vector of the eigenmode of stability matrix for the $\sigma^{th}$ branch; and the symmetry coefficient $\gamma_{\sigma}({\bf k}_{j_{s}})$ determined by normalization condition and geometry. 

{\par}{Previously, the term $e_{i}^{\sigma(k_{j_{s}})}$  in Eq.~\ref{NCompCW} was either assumed or ignored for simplicity in stoichiometric binary cases \cite{Khat1972,Khat1983}. Our DFT+SRO theory directly provides these eigenvectors of the normal concentration modes required to assess ordering behavior of multi-principle element alloys or HEAs at spinodal decomposition temperature. A detailed discussion on model binary alloy is provided  in the supplement Section S1.} 

{\par}The Eq.~\ref{NCompCW} represents the possible competing types of ordered superlattices (symmetry-broken order) that are incipient in the chemical SRO for a fixed Bravais lattice. {``Incipient ordering" indicates the possible low-temperature ordered structure in the presence of specific short-range order at higher-temperatures. For example, SRO results in broad, diffuse x-ray scattering in the regions where super-structure peaks would appear with LRO at lower temperatures.} Each of the anticipated partially- or fully-ordered cells can be then assessed using DFT calculated $E_{form}^{LRO}$  relative to $E_{form}^{dis}$ -- a direct calculation for state of order rather than estimated using the SRO only. We showcase this in the next section using the SRO and its eigenvectors to estimate the competing phases, and, then, we do direct DFT calculation for the given partially LRO to confirm {(and get proper relative energies)}.  As an aside, the CW analysis for small-cell-type ordering shows that only up to an 8-component HEA can order if at very specific compositions.

\section{Results and Discussion}
\begin{figure}[t]
\begin{centering}
\includegraphics[scale=0.32]{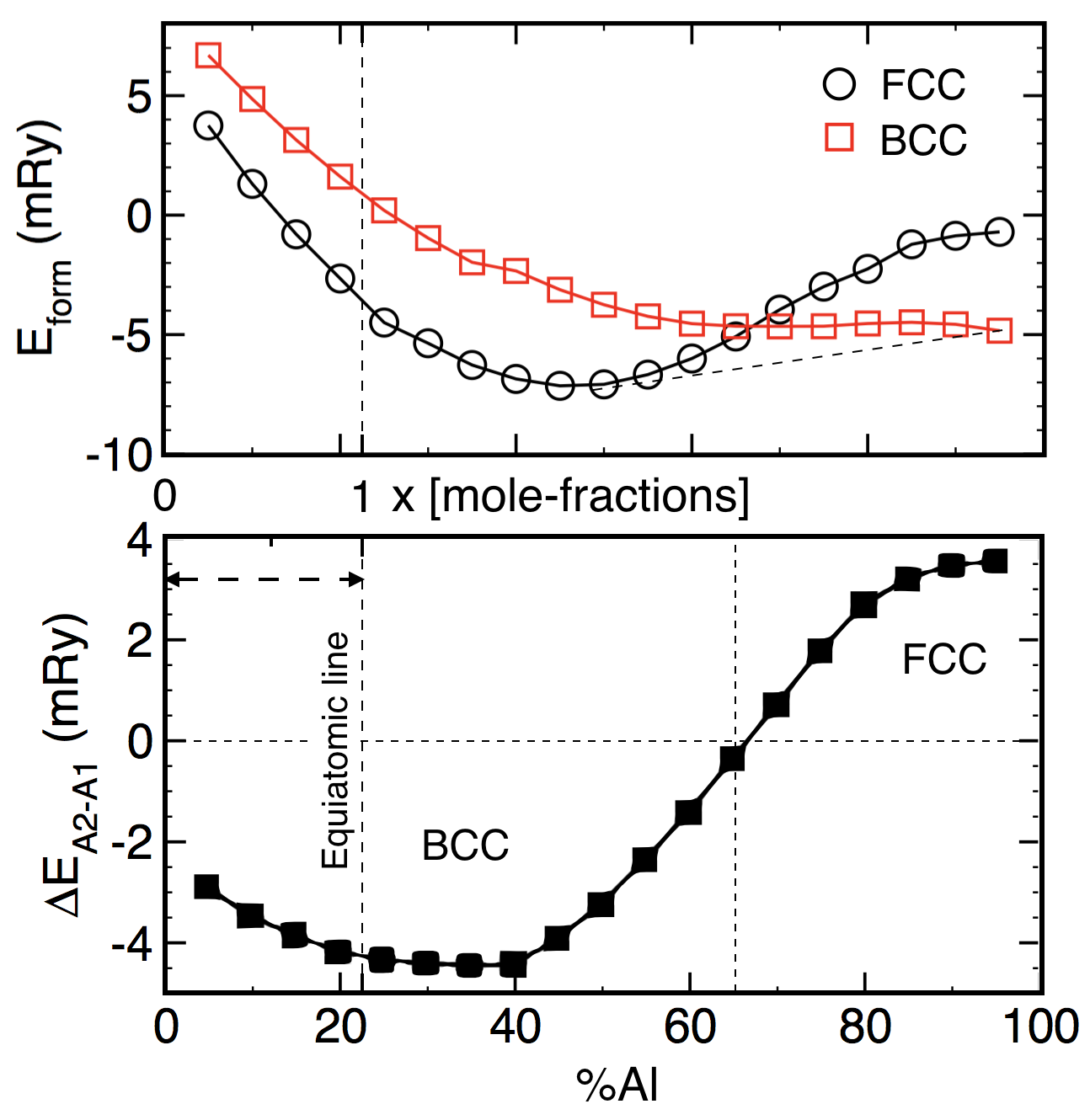}
\caption{(Color online) For Ti$_{0.25}$CrFeNiAl$_{x}$ ($x$ is mole fraction), {the DFT formation energy (top),} and total energy difference (bottom) between A1 (FCC) and A2 (BCC) phases. Experimental Al content ($x \le 1$) is highlighted (arrow). As Ti is fixed,  \%Al $=100x/(3.25+x)$. }
\label{Fig2}
\end{centering}
\end{figure}

 Liu  {\it et.al.}~\cite{L2015} recently reported experimental observations on quinary Ti$_{0.25}$CrFeNiAl$_{x}$ alloys and found BCC as the stable phase throughout the Al composition range of $x = 0-1$ mole fraction (or $0-23.5$ at.\%). 
Our DFT calculated phase stability plot versus Al content, Fig.~\ref{Fig2}, shows BCC as the most stable phase for $x \le 3.25$.
Over the full Al compositions,  FCC is the stable phase from 3.25 mole fraction and beyond.
The initial increase in Al\% further stabilizes the BCC phase (until 40\%Al), i.e., Al plays the role of BCC-phase stabilizer, in agreement with the experiments and CALPHAD~\cite{L2015}.
{To make a further connection, we also show in Fig.~\ref{Fig2} the formation energy (relative to energies of elemental solids in their ground state phases). As is clear, FCC is stable over BCC by 65\%Al, where the alloy then is in a small two-phase region until 100\%Al. }

%%% FIGURE 1

%%% FIGURE BSF
\begin{figure}[t]
\begin{centering}
\includegraphics[scale=0.37]{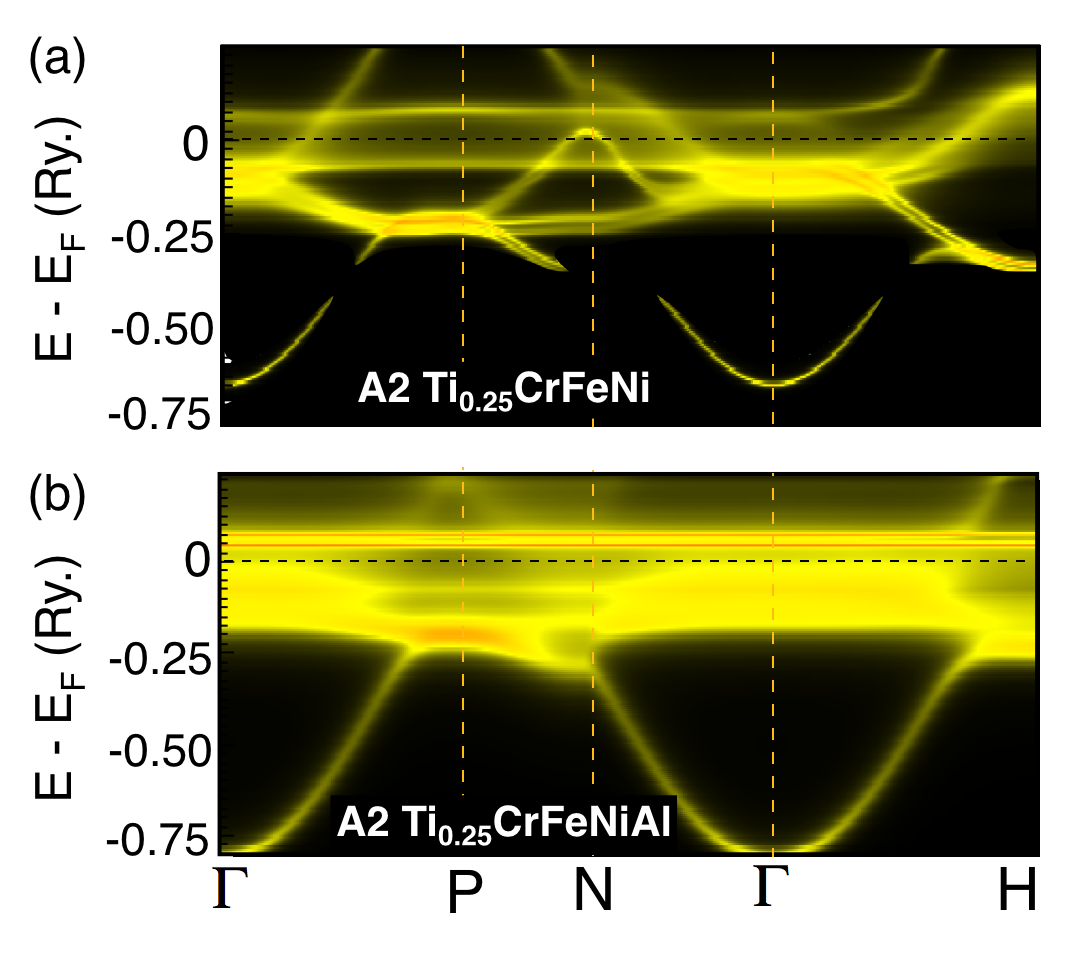}
\caption{(Color online) Block spectral function of  BCC Ti$_{0.25}$CrFeNiAl$_{x}$ at x=0 (a); and x=1.0 (b) along high-symmetry directions of BCC Brillouin zone. Added Al enhances stability of BCC by disorder broadening and enhancing hybridization.}
\label{Fig3}
\end{centering}
\end{figure} 

%Ti$_{0.25}$CrFeNiAli$_{x}$
{\par}To understand the effects of chemical disorder by Al-doping,  we calculate Bloch-spectral function (BSF) of BCC Ti$_{0.25}$CrFeNiAli$_{x}$ using DFT based electronic-structure method for x=0,1 cases, see Fig.~\ref{Fig3}. The BSF is a generalization of order band structure when disorder is present \cite{BSF1,BSF2}. On adding Al to Ti$_{0.25}$CrFeNi, the BSF is smeared out near the Fermi energy (E$_{F}$) due to increased disorder effect. The BSF broadening can directly be related to the inverse of the electron mean free path \cite{Singh2019}. Increased k-space smearing at E$_{F}$ indicates a decrease in electron mean-free path for Fig.~\ref{Fig3}(b) compared to Fig.~\ref{Fig3}(a) -- the shift in dispersion is clearly visible at energies below E$_{F}$. The Al doping enhances disorder and lowers bonding states \cite{SSJ2015}, stabilizing the BCC phase, as also shown in Fig.~\ref{Fig2}. For near equiatomic HEA ($x=1$ Al mole-fraction), the negative formation energy indicates the favorability for the mixing of alloying elements.

{\par} 

{The starting point in our calculations is the high-temperature disorder phase, where entropy contribution mainly arises from the point entropies and disorder local moments (if any). The SRO contribution in completely random state (disorder phase) goes to zero as the configurational entropy is mainly dominated by S$_{mix}$ and S$_{mag}$. The point entropy can be estimated by $S_{pt}= k_B \sum^{N}_{\mu=1} c_{\mu} \ln{c_{\mu}}$ ($-k_B\ln{N}$, where  with $c_{\mu}=1/N$) is a key factor for the formation of single-phase HEAs.} Point entropy increases with increasing number of alloying components, which suppresses the formation of intermetallic phases \cite{Yeh2004,Yeh2006, GNLL2011}, unless enthalpically dominated by favorable chemical interactions amongst pairs of atoms, which grows as $\frac{1}{2} N(N-1) \sim N^{2}$.  So, as $N$ gets larger, enthalpy ($N^{2}$) can win over entropy ($\ln{N}$). Empirically, literature suggests a threshold of $S_{pt} \sim 1.5R$ for an operational definition of high-entropy alloy \cite{M2014}. 
Here, the gas constant $R$ is  $8.314~J{mol}^{-1}K^{-1}$ and $k_B = R/N_{A}$, where $N_{A}$ Avogadro's number.  {For magnetic entropy, we use phenomenological approximation as previously used for paramagnetic Fe-based alloys \cite{PhysRevB.88.085128,PhysRevLett.96.117210, V20063821} with non-integer magnetic moments.}
 
%For example, point entropy of ideal solutions is estimated by S$_{mix}$ =c$_{i}$lnc$_{i}$ ("i" is number of alloying elements), and the magnetic entropy by Smag = ciln(1+?i), where ci is the concentration and ?i the magnetic moment of the ith alloying element [PRB 88, 085128 (2013)].
 
\begin {table}[h]
\begin{centering}
\caption{The chemical, magnetic and total entropy contribution in units of gas constant (R), considering k$_{\rm B}$=1.}
\begin {tabular}{cccccccccccccccccccccc}\hline \hline
x$_{Al}$&&&&&&\multicolumn{5}{c}{Entropy} \\ 
         &&&&  S$_{mix}$  &&& S$_{mag}$   &&&& S$_{Total}$        \\ \hline     
0.25  &&&& 1.40            &&& 0.22  &&&&  1.62     \\ 
0.50  &&&& 1.48            &&& 0.19  &&&&  1.67    \\
0.75  &&&& 1.51            &&& 0.17  &&&&  1.68      \\
1.00  &&&& 1.52            &&& 0.15  &&&&  1.67     \\
\hline \hline
\end {tabular}
\label{tab1}
\end{centering}
\end {table}

{\par}While ${S_{pt}}$ in BCC Ti$_{0.25}$CrFeNiAl$_{x}$ is large, but the magnetic character of alloying elements (Cr/Ni/Fe) also suggests the possible contribution from magnetic entropy. For BCC Ti$_{0.25}$CrFeNiAl$_{x}$, we estimate the chemical entropy by $S_{pt}$ and the magnetic entropy by $\Delta{S_{\rm mag}}=c{_\nu}\ln(1+\mu_{\rm \nu})$ in units of gas constant (R) [setting k$_{B}$=1], here $\mu_{\rm \nu}$ is the magnetic moment of $\nu^{th}$ element. The chemical, magnetic and total entropy contribution are tabulated in Table~\ref{tab1}. Increasing Al content increases the chemical entropy, which saturates at equiatomic Al, whereas we notice slight decrease in the magnetic entropy as Fe losses its magnetic character with increases Al content. 

{\par} To reveal the ordering in Ti$_{0.25}$CrFeNiAl$_{x}$ and find out the plausible reasons for disagreement between experiments and CALPHAD, we chose four sets of alloys, i.e., $x=0.25, 0.50, 0.75, 1.00$, permitting a one-to-one comparison to experimental composition range of Liu  {\it et.al.}~\cite{L2015}. As SRO is dictated mostly by the electronic-structure of the alloy, the origins of the observed ordering tendencies in Ti$_{0.25}$CrFeNiAl$_{x}$ can be determined, i.e., all the competing effects (e.g., band-filling, Fermi-surface nesting, atomic size, and charge transfer) can be assessed \cite{SSJ2015,Singh2018}.

\begin{figure}[t]
\begin{centering}
\includegraphics[scale=0.32]{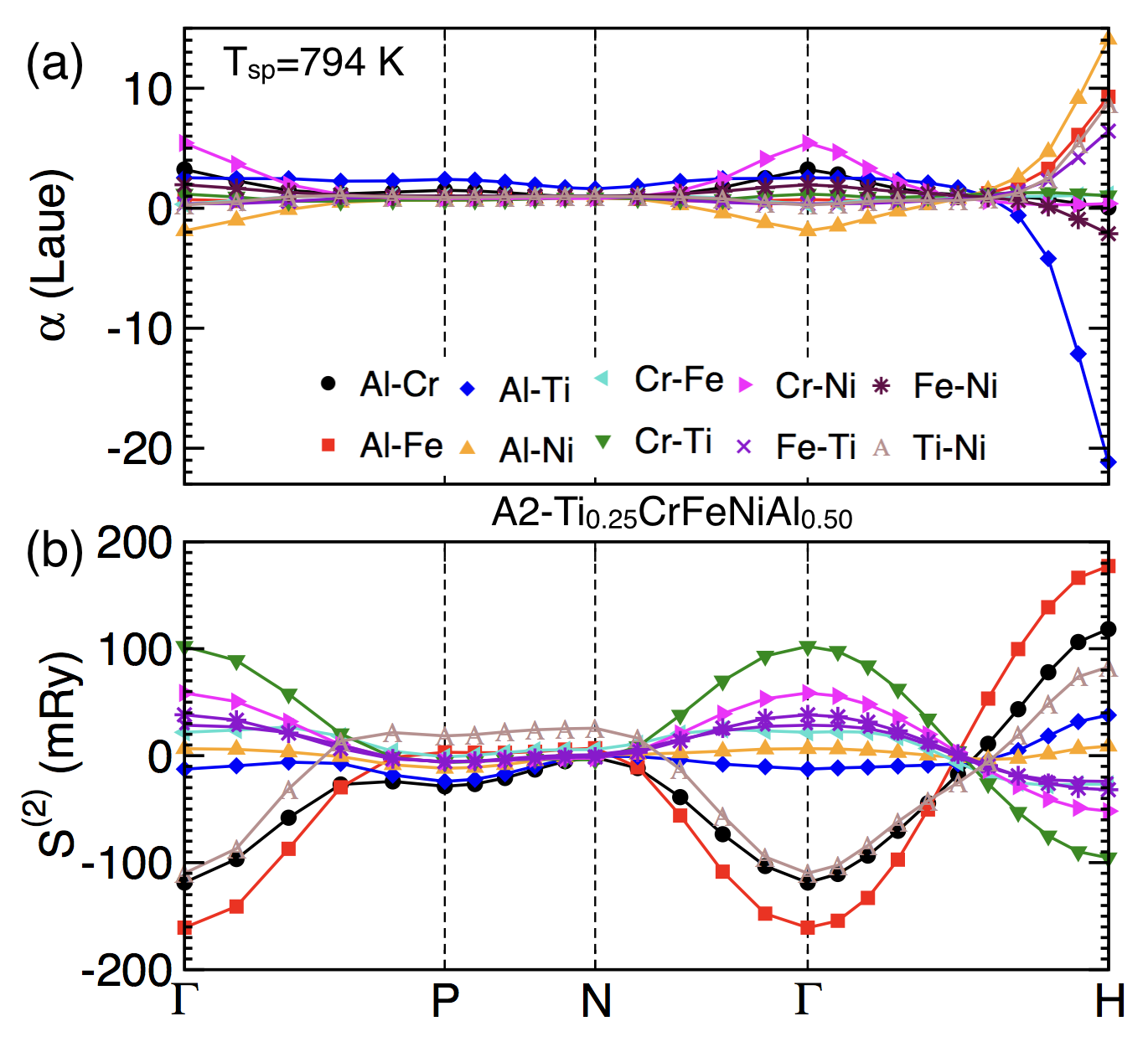}
\caption{(Color online) For BCC Ti$_{0.25}$CrFeNiAl$_{0.50}$, $\alpha_{\mu\nu}({\bf k};T)$ and S$^{(2)}_{\mu\nu}({\bf k};T)$ along Brillouin zone symmetry lines. At T=1.15T$_{sp}$, $\alpha_{Al-Ni}$ in (a) shows dominant SRO with peak at ${\bf k}_{o}\,$=H=\{111\} indicating B2-type ordering, but $S^{(2)}_{Al-Fe}$ in (b) drives the instability.}
\label{Fig4}
\end{centering}
\end{figure}

{\par} We exemplify SRO predictions and concentration-wave analysis on Ti$_{0.25}$CrFeNiAl$_{0.5}$. In Fig.~\ref{Fig4}, we plot the SRO and interchange energies S$^{(2)}_{\mu\nu}$ at 1.15${T}_{{sp}}$ (794 K calculated). As $\alpha^{-1}({\bf H};T)]_{\mu\nu}$ vanishes at ${T}_{{sp}}$, i.e., the  SRO diverges at ${\bf k}_{o}={\bf H}=(111)$, which indicates the B2-type (CsCl) ordering instability. {At ${T}_{{sp}}$, $\alpha_{\mu\nu} ({\bf H})$~has a dominant SRO peak for Al-Ni pair (followed by Al-Fe and Ti-Ni) in Fig.~\ref{Fig4}(a), whereas the instability in S$^{(2)}$ is driven by  Al-Fe pairs (followed by Al-Cr) in Fig.~\ref{Fig4}(b).} 
This odd (but correct) result occurs due the probability sum rule of $\alpha_{\mu\nu}({\bf k})$, and its inverse-relation with S$^{(2)}_{\mu\nu}({\bf k})$ \cite{AJP1995,DDJ2012,SSJ2015,Singh2018,Singh2019}. The instability at {\bf H} in BCC-Ti$_{0.25}$CrFeNiAl$_{0.50}$ occurs when (at least) one of the eigenvalues of the correlation matrix is maximum, i.e., inverse of the corresponding pair-correlation component of the correlation matrix vanishes. The relative polarization of concentration waves is represented by the eigenvector corresponding to the vanishing eigenvalue of aforementioned correlation matrix in the Gibbs space \cite{SSJ2015, AJPS1996, Fontaine1979, BKK1969}. We extract the eigenvector corresponding to {\bf H} and {\bf P} for Ti$_{0.25}$CrFeNiAl$_{0.50}$ at ${T}_{{sp}}$ to analyze the B2 and L2$_{1}$ type ordering. The eigenvectors at  T$_{sp}$ helps estimate the LRO parameters to solve the Eq.~\ref{NCompCW} for occupation probabilities \cite{SSJ2015}.

%  FIGURE  SRO

{\par} Using disordered state information of elemental composition, structure factor, instability vector and eigenvector (at ${T}_{{sp}}$), the concentration-wave (probability $n({\bf r})$) for BCC-Ti$_{0.25}$CrFeNiAl$_{0.50}$ can be written as (using Ni as the `host' arbitrarily):
\begin{equation}
\begin{split}
\small{
%n({\bf r}) = 
\left[\begin{matrix} { n^{Al}({\bf r})} \\   {n^{Cr}({\bf r})} \\ {n^{Fe}({\bf r})} \\ {n^{Ti}({\bf r})} \end{matrix}\right]
=  \left[\begin{matrix} 0.133 \\ 0.269 \\0.269 \\0.050 \end{matrix} \right]  
+  \frac{\eta_{B2}}{2}\left[\begin{matrix} +1.167 \\ +0.027 \\ -0.521 \\ +0.271 \end{matrix}\right]  e^{i(111)\cdot{\bf r}}   .
}
\end{split}
\label{B2}
\end{equation}
where $\eta_{B2}$ is the LRO parameter for B2-order, which is used for occupation probability determination for B2 sublattices at $T_{sp}$, i.e., ${\bf{a}}$=$(000)$ or ${\bf{b}}$=$(\frac{1}{2}\frac{1}{2}\frac{1}{2})$. With sum rules ($\sum_{\alpha=1}^{N} c_\alpha = \sum_{\alpha=1}^{N} n_\alpha({\bf r}) =1$),  $n^{Ni}({\bf r})$ is obtained.

{\par} Upon ordering the A2 (BCC) lattice splits into two simple-cubic sublattices with $\{111\}$ ordering vector. 
Here, the maximum possible LRO corresponds to Al at sublattice $(\frac{1}{2}\frac{1}{2}\frac{1}{2})$, i.e., occupation probability of Al  vanishes ($n^{Al}=0$), and Eq.~\ref{B2} (right side for Al) simplifies to  $0.133 - 0.5\times{1.167}\times{\eta_{B2}}=0$. 
As the alloy cannot have negative probabilities, the maximum LRO parameter is $\eta_{B2}=0.22794$ and corresponds to symmetry-breaking of ${\bf k}_{0}=\{111\}$ at $T=T_{sp}$. 
The occupation probabilities $n({\bf r})$ at sub-lattices $(0,0,0)$ and $(\frac{1}{2}\frac{1}{2}\frac{1}{2})$ calculated using the maximum allowable LRO are  (0.26600, 0.27208, 0.20962, 0.08089, 0.17141) and (0.26592, 0.32838, 0.01911, 0.38659), respectively. The calculated occupations probabilities from first-principles SRO theory shows that in symmetry-breaking process, i.e., on ordering Al breaks the symmetry of BCC lattice and preferentially occupies the cube corner. 
The Al probability vanishes at one of sublattice, i.e., at $(\frac{1}{2}\frac{1}{2}\frac{1}{2})$. This way Al$_{0.50}$CrFeNiTi$_{0.25}$ partially orders into B2 superstructure, where partially-ordered B2 phase has lower energy than A2 phase,
where the energy gain from partial ordering to this state ${\Delta E^{\rm B2-A2}}$  is given by the energy difference of BCC and partially-ordered B2 (site probabilities given by Eq.~\ref{B2}), i.e., ${\Delta E^{\rm B2-A2}} = {E^{\rm B2} - E^{\rm A2}} = -3.48$~mRy.  The partially-ordered  energetics is very sensitive to order parameter and sublattice occupations. Thus, care must be taken in calculating occupation probabilities such that sum rules are obeyed. The  B2 phase is stabilized with respect to A2 phase using similar calculations for $x=0.75$ and $1$, yielding ${\Delta E^{\rm B2-A2}}$ of -2.98 mRy and -8.33 mRy, respectively.

\begin{figure}[t]
\begin{centering}
\includegraphics[scale=0.33]{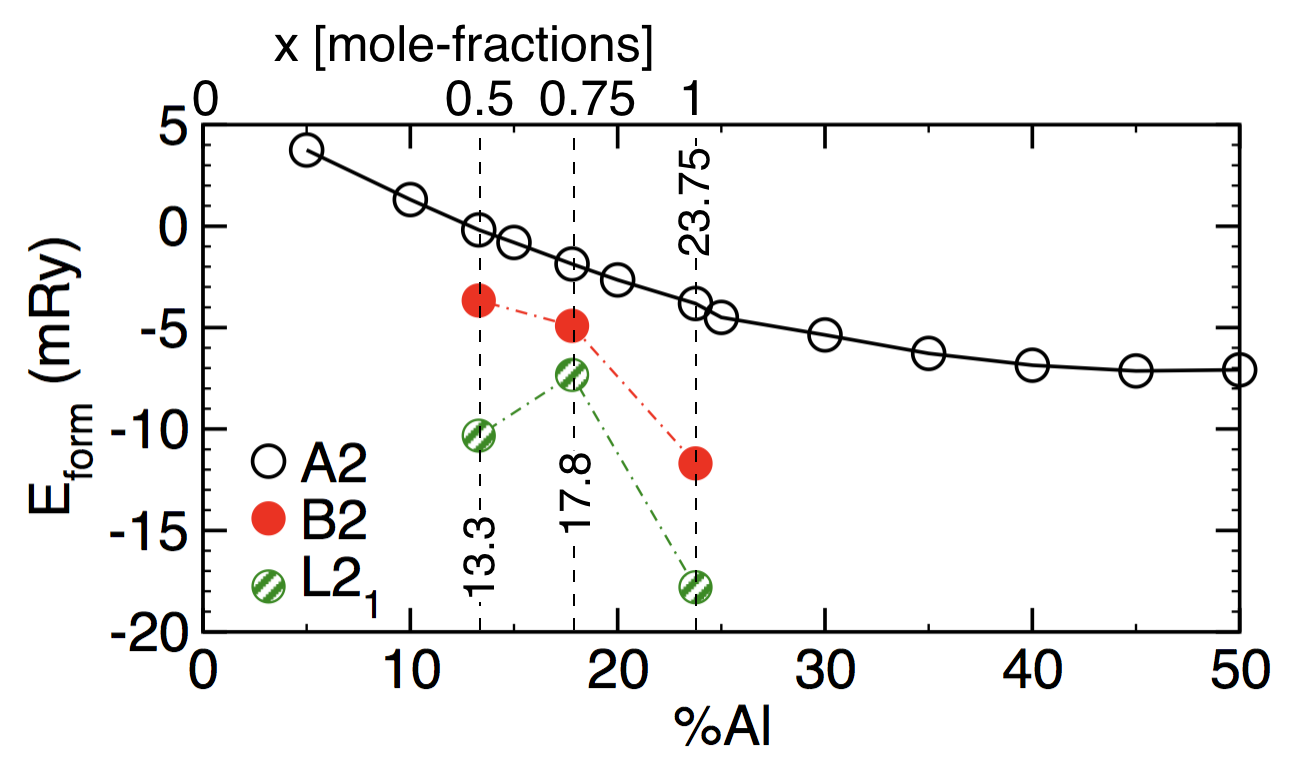} 
\caption{{(Color online) For Ti$_{0.25}$CrFeNiAl$_{x}$ ($x$ is mole fraction and  \%Al $=100x/(3.25+x)$), the DFT E$_{form}$ of A2 and of partially-ordered  {B2} and {L$2_1$} at $x=0.5, 0.75, 1.0$ (where E$^{A2}_{form}$ are $-0.19$, $-1.87$, and $-3.81$~mRy, resp.). The lowest A2 value ($-7.14$~mRy) is at 45\%Al.  E$^{B2-A2}$ values are reported in Table~\ref{tab2}.}}
\label{Fig5}
\end{centering}
\end{figure}

{\par}As already discussed, the A2 phase can order into different lower-symmetry structures in going from high-temperature (disorder) phase to lower-temperature (partially-ordered) phases. {At first, on lowering temperature, the A2 lattice shows regions of B2 superstructure by lowering symmetry along $\{111\}$. Upon further lowering the temperature, symmetry can break via  $\{\frac{1}{2}\frac{1}{2}\frac{1}{2}\}$ (secondary ordering) and may order into L$2_{1}$ superstructure depending on the material characteristics.  Relative to A2, we show E$_{form}$ for partially-ordered B2 and L$2_{1}$ phases in Fig.~\ref{Fig5}, where site occupations for the partially-ordered phase are obtained from the SRO+CW approach, as discussed in Eq.~\ref{B2}\&~\ref{L21} (also see Section S1 of supplement).  As can be seen in Fig.~\ref{Fig5}, there is significant gain in energy from A2$\rightarrow$B2 at $x=1$ (23.75\%Al), where B2 has still has significant (temperature-dependent) point entropy given by the occupation variables of the partially-ordered B2 state. }

%%% FIGURE 4 (NEW)

{\par}The A2-to-B2 (2$^{nd}$-order) transition is defined by one LRO parameter, while L2$_{1}$ requires two: $\eta_{1}$ and $\eta_{2}$. Our linear-response predicts SRO, which yields correct second-order transition, which often dictates expected first-order phase-transitions. {However, first-order transitions have a discontinuous $\eta(T)$ versus T that requires {\it ab-inito} thermodynamic simulation to predict it directly.} As the L2$_{1}$ phase occurs due to secondary-ordering, a qualitative prediction can be made if we deliberately break the symmetry of Ti$_{0.25}$CrFeniAl$_{0.50}$ by ordering formed by ${\bf k}_{1}=\{111\}$ wavevectors (B2 ordering) and ${\bf k}_{2}=\{\frac{1}{2}\frac{1}{2}\frac{1}{2}\}$ (Heusler or  L2$_{1}$ ordering)  using the eigenvectors estimated from the primary transition, i.e., in A2-B2 at $T=T_{sp}$. $L2_{1}$ is then  represented by a concentration-wave with vector site probabilities $n({\bf r})$
\begin{equation}\label{L21}
\small{
\begin{split}
\left[\begin{matrix} { n^{Al}({\bf r})} \\   {n^{Cr}({\bf r})} \\ {n^{Fe}({\bf r})} \\ {n^{Ti}({\bf r})} \end{matrix}\right]
=  \left[\begin{matrix} 0.2375 \\ 0.2375 \\0.2375 \\0.0500 \end{matrix} \right]  
   +    \left[\begin{matrix} +1.167 \\ +0.027 \\ -0.521 \\ +0.271 \end{matrix}\right]  \times 
     \lbrace   \frac{\eta_{1}}{4} e^{2{\pi}i{\bf k}_{1}\cdot{\bf r}}  \\
 +  \frac{\eta_{2}}{2}   \left[  \cos\left\lbrace2{\pi}{\bf k}_{2}\cdot{\bf r}\right\rbrace   +   \sin\left\lbrace2{\pi}{\bf k}_{2}\cdot{\bf r}\right\rbrace  \right]  \rbrace   . 
\end{split}
}
\end{equation}
{The last term in Eq.~\ref{L21} reflects the additional term over B2 in Eq.~\ref{B2}, which is an enriching minority components along $\{\frac{1}{2}\frac{1}{2}\frac{1}{2}\}$ for L2$_{1}$ ordering.}  

{\par}  It is convenient to describe A2 lattice with L2$_{1}$ ordering with the help of the four interpenetrating A1 sublattices: $(000)$, $(\frac{1}{2}\frac{1}{2}\frac{1}{2})$, $(\frac{1}{4}\frac{1}{4}\frac{1}{4})$ and $(\frac{3}{4}\frac{3}{4}\frac{3}{4})$, with twice the A2 cubic lattice parameter. The maximum LRO $\eta_{1}$ and $\eta_{2}$ correspond to the sublattice $(\frac{1}{2}\frac{1}{2}\frac{1}{2})$ and $(\frac{1}{4}\frac{1}{4}\frac{1}{4})$ for which `Al' site probability vanishes first, i.e., $n^{Al}=0$ and  $0.133-\frac{1}{4}\times1.167 \times \eta_{1}(T)=0$ and $0.133- \frac{1}{4} \times1.167 \times \eta_{2}(T)=0$ at $T=T_{sp}$. The resulting LRO parameters are $\eta_{1}=0.45587$ and $\eta_{2}=0.32235$ for sub-lattices $(\frac{1}{2}\frac{1}{2}\frac{1}{2})$ and $(\frac{1}{4} \frac{1}{4}\frac{1}{4})$, respectively, at $T=T_{sp}$ (structural files for A1/A2/B2/L2$_{1}$ used for total energy calculations along with computed occupation probabilities (Table S1) and  lattice-constants (Table S2) are provided in the supplement Section S2).

{\par} {For non-stoichiometric cases of Ti$_{0.25}$CrFeNiAl$_{x}$, SRO predicts partially-ordered B2 and L2$_{1}$ state (see compositions inTable S1). The negative energy gain at x=0.50 of partially ordered B2 and L2$_{1}$ phases  with respect to A2 phase is $E^{\rm B2}-E^{\rm A2}=-3.48$ mRy and ~$E^{\rm L2_{1}}-E^{\rm A2}=-10.14$ mRy, respectively.} 
Similar to x=0.50, Ti$_{0.25}$CrFeNiAl$_{x}$ show thermodynamically stable B2 and L2$_{1}$ phases with respect to A2 at x =0.75 and x= 1.0.  The ordering energy difference for B2 and L2$_{1}$ phases with respect to disorder phase at x=0.75 and x=1.0 is ($E^{\rm B2}-E^{\rm A2} = -2.98$~mRy  and  $E^{\rm L2_{1}}-E^{\rm A2} =-5.47$~mRy for $x=0.75$) and ($E^{\rm B2}-E^{\rm A2} =-8.33$~mRy and  $E^{\rm L2_{1}}-E^{\rm A2}=-14.44$~mRy), respectively. 
The estimated spinodal temperature for Ti$_{0.25}$CrFeNiAl$_{x}$ is $T_{\rm sp}=(794; 1802; 1190)$~K at $x=(0.50, 0.75; 1.00)$. %The increase in Al content increases  stability with disorder, i.e., disorder phase remains stable over larger temperature range. 

{\par}The instability in alloy occur at the spinodal temperature above which mixture remains homogeneous. {For homogeneous fluctuations from linear-response approach, we can write the change of free-energy in terms of concentration fluctuations and pair-correlation function evaluated at the point of instability.} The estimated change in energy, $\delta{E^{\rm X-A2}}$, can be written as \cite{DDJ2012,SSJ2015}
\begin{equation}
\small{
\delta{E^{X-A2}} =\frac{1}{2}\sum_{j_{s}}\sum_{\alpha\ne{\beta}}S^{(2)}_{\alpha\beta}({\bf k}_{j_{s}}; T) \delta{c^{\dagger}_{\alpha}}({\bf k}_{j_{s}}) \delta{c_{\beta}}({\bf k}_{j_{s}}) ,
} 
\end{equation}
where X is superlattice order with instability in ${\bf k}_{j_{s}}$  and associated concentration changes, $\delta{c_{\alpha}}({\bf k}_{j_{s}})$. This `back-of-the-envelope' calculation uses the pair-interchange energies to estimate directly the energy gain for particular ordering without additional calculations. In Table~\ref{tab2}, the B2 energy gain estimated from SRO, $\delta{E^{\rm B2-A2}}$ (${\bf H}$) for $x=(0.50; 0.75; 1.00)$, shows the same trend as the direct DFT calculations. Importantly, SRO estimate gives robust trends without additional calculations.  {The direct evaluation of the formation energies, $E_{form}$, of the partially LRO state relative to the disordered state establish order-disorder transition temperatures, for example, $k_B T_{od} = E_{form}^{LRO} - E_{form}^{dis}$ in ordering systems \cite{AKJ2010,ZJ2019} and slightly more complicated in segregating systems \cite{ZTJ2007}.} 
%Temperatures are thermodynamically consistent $T_{sp} \le T_{od}$ if evaluated correctly.

\begin {table}[b]
\begin{centering}
\caption{For Ti$_{0.25}$CrFeNiAl$_{x}$ we show the B2-A2 energy difference from SRO ($\delta E_{SRO}$) \cite{SSJ2015}, which compares well with a direct DFT calculated energy difference ($\Delta{E}$) using the ASA or a better integration over VP for charges \cite{MECCA}.}
\begin {tabular}{cccccccccccccccccccccc} \hline \hline 
$x$    &&&& \multicolumn{5}{c}{$\Delta E^{\rm B2-A2}$ (mRy) } \\ 
         &&& $ \delta E_{SRO}$  &&& $\Delta{E_{ASA}}$   &&& $\Delta{E_{VP}}$        \\ \hline     
0.50  &&& -5.03            &&& -3.48  &&&  -3.26     \\
0.75  &&& -4.23            &&& -2.98  &&&  -2.70      \\
1.00  &&& -9.43            &&& -8.33  &&&  -7.88      \\
\hline \hline
\end {tabular}
\label{tab2}
\end{centering}
\end {table}

{\par} The difference between direct DFT calculations and that from the SRO is easily understood.  Upon breaking symmetry with a LRO parameter into B2, the site charges and dispersion change directly responding to the new symmetry and changes in electronic charge density, whereas using the SRO from linear-response, the dispersion is fixed to the random alloy -- hence, it is computational less expensive and provide fast estimate -- albeit a good one, especially for trends.  One may consider using this for fast evaluation of HEA before spending time on more accurate calculations. The SRO can be viewed in Gibbs' composition space (a Barycentric coordinate system) and the SRO can be expanded like in finite-element codes, permitting scans of composition space with but a few compositions initially tested.  

{\par} Our calculations of Ti$_{0.25}$CrFeNiAl$_{x}$ predict stable B2 (Al-poor region) and B2+L2$_{1}$  (Al-rich region), whereas experiments did not find any signature of L2$_{1}$ ordering. In contrast, the CALPHAD phase diagram suggests A2+B2+L2$_{1}$ phases, i.e., a mixed ordering transformation during slow cooling process \cite{L2015}. {In some cases, these LRO can happen at much lower temperatures as the kinetics (diffusion) is limited and alloy does not transform due to larger energy barrier, unless assisted some way, such as by shear. This has been observed in FCC to HCP transitions at very low temperature for CrMnFeCo \cite{HE2019437}. It could also happen that SRO is simply frozen (quenched disorder), where experiments from high-T samples are characterized at room temperature \cite {MARUCCO1994267}.}

%%% FIGURE 5
\begin{figure}[t]
\begin{centering}
\includegraphics[scale=0.34]{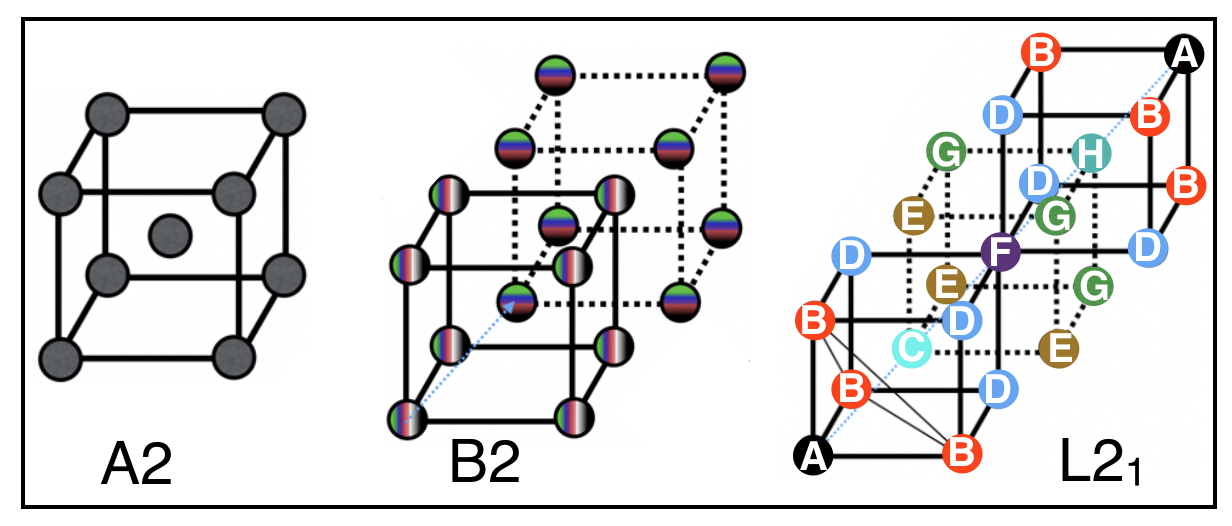}  
\caption{(Color online) Schematic unit cells of high-symmetry (disorder) A2 HEA, which, upon cooling, lowers symmetry to a partially-ordered superstructure, e.g., B2  and L2$_{1}$. Sites: 1~A (black), 3~B (red), 1~C (yellow), 3~D (blue), 3 E (brown), 1~F (purple), 3~G (green) and 1~F (orange), i.e.,  8 sites (16 atoms) reference to $2\times2\times2$ A2 cubic cell. Partially-ordered L2$_{1}$ (predominantly A on A sites, B on B sites, and so on) can only be fully-ordered at perfect stoichiometry. }
\label{Fig1}
\end{centering}
\end{figure}

{\par} In Fig.~\ref{Fig1}, we show that there are 8 possible distinct sites in L2$_{1}$ superstructure that can be populated without destroying cubic symmetry, i.e., 2$\pi$/3 rotation along $\left\langle{111}\right\rangle$, and mirror symmetry along $\left\langle{110}\right\rangle$. An HEA with more than 8-components, even with stoichiometric compositions, cannot be populated with small-cell cubic order. Thus, we can infer from Fig.~\ref{Fig1} that L2$_{1}$ order in quinary systems can exist. So, HEAs often exhibit a series of ordering transitions with one or more partially-ordered phases, and it ultimately reaches to fully-ordered states (if stoichiometric). In general, upon cooling,  A2 goes to  B2-type order by lowering its symmetry along $ \left\langle{111}\right\rangle$, e.g., in a binary A2 to B2 with corner (center) sites are perfectly ordered.  In a HEA, as temperature is lowered, the B2 phase can only be partially-ordered, then B2 phase breaks symmetry along $\left\langle\frac{1}{2}\frac{1}{2}\frac{1}{2}\right\rangle$ to a Heusler-type superstructures, e.g., DO$_{3}$ (AB$_{3}$) or L2$_{1}$ (ABC$_{2}$), {typically partially-ordered}. 

{For clarity, we note that linear-response can simultaneously find all ordering wavevectors ({\bf k}) associated with specific underlying Bravais lattice (e.g., A1/A2/A3), i.e., there is no restriction on SRO or LRO modes. All possible arbitrary length correlations, which is impossible in real-space methods, indeed can provide full real-space details when inverse Fourier transformed \cite{Singh2018}. Notably, not all the ``incipient LRO" leads to LRO states, e.g., a superstructure usually forms at certain stoichiometric compositions. Therefore, the degree of LRO decreases as the compositions deviates from stoichiometry. This deviation from the stoichiometry makes the transformation kinetics slower, which may need thousands of hours of aging time \cite {MARUCCO1994267}, but SRO in the high-T disordered phase always can occur.  Considering  our work is based on full electronic structure (chemical and magnetic), we are also able to capture the effects of hybridization, Fermi-surface, band-filling, Kohn anomalies, etc., on the SRO and directly ``fingerprint'' the origin for ordering \cite{SSJ2015,Singh2018,Singh2018_2,Singh2019}.}

\section{Conclusion}
{For arbitrary HEAs, we presented a concentration-wave (CW) analysis of results from electronic-structure-based thermodynamic linear-response theory (SRO) for a fast assessment of all possible low-temperature competing partially-ordered states. The SRO, arising from the electronic-structure (dispersion and energetics) of the disordered solid-solution at specific concentrations, gives detailed information on the unstable ordering modes inherent in the HEA, as well as their electronic origins. A CW analysis directly identifies partially- or fully-ordered unit cells for competing states as well as their sub-lattice occupations. We can then use these identified partially-ordered unit cells in a direct DFT calculation to obtain quantitative results for energy difference between all competing states (or relative to the  fully-disordered solid-solution with a given underlying Bravais lattice. The estimate of energy gain for each structure with the relevant competing states, identified from SRO, can also be done from the SRO parameters using a `back-of-the-envelope' estimate, which provides a good qualitative trend. }

{\par}{The DFT+SRO assures the necessary symmetry condition for the thermodynamic stability of the ordered superstructure, and SRO+CW predicts the atomic arrangement of the partially ordered or ordered superstructure. The proposed analysis method also rejects any a priori assumptions on the crystal structure of ordered phases.} 
We exemplified the stability, electronic-dispersion, and the Warren-Cowley SRO parameter for the case of the HEA Ti$_{0.25}$CrFeNiAl$_{x}$. 
Our direct DFT calculations of total-energy shows that increasing Al stabilizes BCC phase, and FCC phase becomes stable above  above \%65-Al. 
The DFT-based SRO calculations with a concentration-wave analysis predicts competing B2 and L2$_{1}$ phases in the Al-rich region, which is in good agreement with CALPHAD study of Liu  {\it et.al.}~\cite{L2015} that has competing B2 and L2$_{1}$ phases in the Al rich-region (0.75 to 1 mole fraction). 
{The uniqueness of our approach lies in the fact that it provides fast, electronic-structure-based assessment of thermodynamic stability and ordering in multi-principal-element alloys, in particular without need for large unit-cell  to model disorder, which will accelerate the design of new systems.}

\section{Acknowledgements}
The work was supported by the U.S. Department of Energy (DOE), Office of Science, Basic Energy Sciences, Materials Science and Engineering Division. The research was performed at Iowa State University and the Ames Laboratory, which is operated for the U.S. DOE by Iowa State University under contract No. DE-AC02-07CH11358. AA acknowledges funding from the National Center for Photovoltaic Research and Education (NCPRE) funded by Ministry of New Renewable Energy (MNRE), Government of India and IIT Bombay.

\biboptions{sort&compress}
\bibliography{Manuscript1}
\bibliographystyle{rsc}
\end{document}